\documentclass[aps,prl,twocolumn]{revtex4}
\usepackage{amsmath}
\usepackage{graphicx}
\begin{document}

\title{Scattering Theory of Charge-Current Induced Magnetization Dynamics}

\author{Kjetil Magne D{\o}rheim Hals}
\author{Arne Brataas}
\affiliation{Department of Physics, Norwegian University of Science and Technology, NO-7491, Trondheim, Norway}
\author{Yaroslav Tserkovnyak}
\affiliation{Department of Physics and Astronomy, University of California, Los Angeles, California 900095, USA}

%%%%%%%%%%%%%%%%%%%%%%%%%%%%%%%%%%%%%%%%%%%%%%%%%%%%%%%%%%%%%%%%%%%%%%%%%%%%%%%
\begin{abstract}
In ferromagnets, charge currents can excite magnons via the spin-orbit coupling. We develop a novel and general scattering theory of charge current induced macrospin magnetization torques in normal metal$|$ferromagnet$|$normal metal layers. We apply the formalism to a dirty GaAs$|$(Ga,Mn)As$|$GaAs system. By computing the charge current induced magnetization torques and solving the Landau-Lifshitz-Gilbert equation, we find magnetization switching  for current densities as low as $ 5\times 10^{6}$~A/cm$^2$. Our results are in agreement with a recent experimental observation of charge-current induced magnetization switching in (Ga,Mn)As.
\end{abstract}

%%%%%%%%%%%%%%%%%%%%%%%%%%%%%%%%%%%%%%%%%%%%%%%%%%%%%%%%%%%%%%%%%%%%%%%%%%%%%%%

\maketitle

%%% New commands %%%
\newcommand{\eq}  {  \! = \!  }
\newcommand{\keq} {\!\! = \!\!}
\newcommand{\kadd}{  \! + \!  }
\newcommand{\ksim}{\! \sim \!}

%%%%%%%%%%%%%%%%%%%%%%%%%%%%%%%%%%%%%%%%%%%%%%%%%%%%%%%%%%%%%%%%%%%%%%%%%%%%%%%
%\section{Introduction}
%%%%%%%%%%%%%%%%%%%%%%%%%%%%%%%%%%%%%%%%%%%%%%%%%%%%%%%%%%%%%%%%%%%%%%%%%%%%%%%

Magnetic random access memories are read using the principle of the giant magnetoresistance effect. Information is written by  non-local magnetic fields, and this limits the density of the memory cells. In small systems, a promising direct and local writing technique is to utilize spin-transfer torques from spin-polarized currents in non-collinear magnetic structures~\cite{SpinTransferTorqueReview}. The interplay between out-of-equilibrium quasi-particle spin-polarized currents and collective magnetization dynamics exhibits a rich variety of phenomena, and receives currently much attention from the scientific community. When a spin polarized current enters a ferromagnet, the itinerant flow of angular momentum normal to the magnetization direction is absorbed. The absorbed angular momentum is transferred to the ferromagnetic order parameter as a torque on the magnetization~\cite{SpinTransferTorqueReview}. Magnetization switching has been seen in several experiments at current densities of the order $10^{6}-10^{8}$~A/cm$^2$ \cite{SpinTransferTorqueReview}. 

Most theories of spin-transfer torques in layered systems disregard spin-orbit interaction in the band structure and impurity potentials. Interestingly, the spin-orbit interaction, which is responsible for spin memory loss, also enables transfer of orbital angular momentum to the collective magnetization. Thus, an unpolarized current can induce a torque on the magnetization even in magnetic homogeneous systems. This has recently been demonstrated theoretically for a simplified magnetic 2DEG Rashba Hamiltonian~\cite{Manchon:prb08} and experimentally seen in (Ga,Mn)As~\cite{Chernyshov:naturephys09}. The experimental results call for 1) an extension of current induced torque theories to include spin-orbit interaction which is valid for general band-structures  and disorder as well as a 2) quantitative computation of the current induced magnetization dynamics. 

We consider a ferromagnet in electric contact with normal metals. The torque on a macrospin ferromagnet characterized by a spatially independent unit vector $\mathbf{m}$ can be written as 
\begin{equation}
\mathbf{\dot{m}} = -\gamma \mathbf{m}   \times \left[   \mathbf{H}_\text{eff} (\mathbf{m}) + \mathbf{H}_s ( \mathbf{m} )  + \mathbf{H}_c (\mathbf{m}) \right] \, ,
\end{equation}% 
where  $\gamma$ is the gyromagnetic ratio and $\mathbf{H}_\text{eff}$ is the effective magnetic field arising from the ferromagnet's equilibrium energy. $\mathbf{H}_s$ and $\mathbf{H}_c$ are induced by non-equilibrium quasi-particle transport. 
$\mathbf{H}_s \propto \boldsymbol{\mu}_s$ describes the spin-transfer torque induced by a spin accumulation $\boldsymbol{\mu}_s$ in the normal metals adjacent to the ferromagnet. Our main focus is the charge-current induced torque, represented by $\mathbf{H}_c \propto V$, induced by a voltage bias $V$ across the ferromagnet without any spin bias.  When the spin-orbit interaction vanishes so does $\mathbf{H}_c$.

We will in this paper derive expressions for the field $\mathbf{H}_c$ valid for general bandstructures and disorder, and with that formulate a novel and general theory for  current-induced magnetization dynamics in layered N$|$F$|$N structures. The theory describes how unpolarized charge currents can excite magnons in the presence of spin-orbit coupling. We also find generalized expressions for $\mathbf{H}_s$ that is valid even when there is spin-orbit coupling and the spin-transfer torques cannot just be evaluated from the loss of transverse spin current. Without spin-orbit coupling, $\mathbf{H}_c=0$ and our formalism reduces to the magnetoelectronic circuit theory~\cite{Brataas:EurPhysJB01}. We apply our theory to a disordered GaAs$|$(Ga,Mn)As$|$GaAs system, and find that an unpolarized charge current induces magnetization reversal for current densities as low as $5\times 10^{6}$~A/cm$^2$.  This is in agreement with recent experimental results and demonstrates the usefulness and potential of the scattering theory. 

Ferromagnetic semiconductors are examples of systems with a large intrinsic spin-orbit coupling which opens new paths to the manipulation of the magnetization direction. In (Ga,Mn)As, ferromagnetism is mediated by holes, and the magnetic anisotropy energy depends on the hole concentration. This together with the ability to control the hole density by gate voltages in semiconductors, enables magnetization switching~\cite{Chiba:nature08}. In contrast, in our case, it is the current induced out-of-equilibrium hole population that initiates magnetization switching. 

%%%%%%%%%%%%%%%%%%%%%%%%%%%%%%%%%%%%%%%%%%%%%%%%%%%%%%%%%%%%%%%%%%%%%%%%%%%%%%%
%\section{Theory}
%%%%%%%%%%%%%%%%%%%%%%%%%%%%%%%%%%%%%%%%%%%%%%%%%%%%%%%%%%%%%%%%%%%%%%%%%%%%%%%
\begin{figure}[ht]
\centering
\includegraphics[scale=0.75]{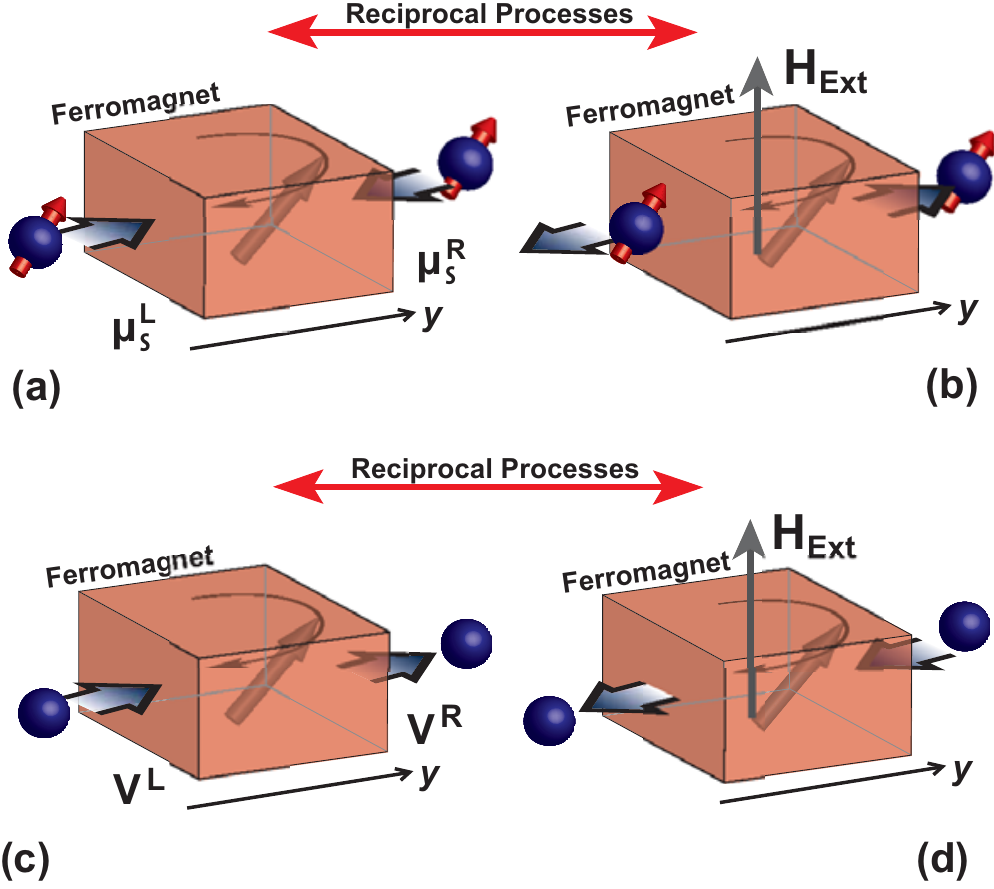}
\caption{(Color online) (a) and (b) are reciprocal spin processes: (a) spin transfer torque induced magnetization dynamics by a spin bias $\boldsymbol{\mu}_s^L$ ($\boldsymbol{\mu}_s^R$) in the left (right) reservoir, (b) spin pumping by  a precessing magnetization into the left and right reservoir. (c)-(d) are reciprocal charge processes, (c) magnetization dynamics induced by voltages $V_L$ and $V_R$, (d) charge pumping by precessing magnetization between the left and right reservoir. Observation of one of the processes implies, by Onsager reciprocal relations, the existence of the reciprocal process. }
\label{fig:ReciprocalProcesses}
\end{figure} 

Spin is not conserved in systems with spin-orbit coupling so spin transfer torques cannot be found from the discontinuity of the transverse spin-current at normal metal$|$ferromagnet interfaces. Instead, torques can be formulated in terms of the spatially dependent exchange-correlation energy~\cite{Nunez:SolidStateComm06}, but this requires a computation of the out-of-equilibrium spin density in the ferromagnet. We choose another route using the Onsager reciprocal relations resulting in a compact result for the torques in terms of the scattering matrix of the complete system without requiring a detailed knowledge of the possibly complicated spatially dependent out-of-equilibrium spin density. 

The derivation of our formalism is based on the Onsager reciprocity relations, which state the following~\cite{de_Groot:bok}: \textit{Assume the system is described by the parameters $\left\{ q_i|i=1,\ldots,N   \right\}$, and the rate of change $\dot{q}_i$ is induced by the thermodynamic force $X_i$. In linear response $\dot{q}_i=\sum_{j=1}^{N} L_{ij}X_j$. Then $L_{ij}= \epsilon_i\epsilon_jL_{ji}$, where $\epsilon_i=1$ ($\epsilon_i=-1$) if $q_i$ is even (odd) under time reversal.}  In a N$|$F$|$N system a voltage  $V^{L(R)}$  and a spin accumulation $\boldsymbol{\mu}_s^{L(R)}$  at the left (right) node  induce a rate of change in the charge and spin density in the left (right) reservoir, which by the continuity relation equals the charge and spin flows $I^{L(R)}$  and $\mathbf{I}_s^{L(R)}$ into the ferromagnet. $\mathbf{X}_c\equiv\left( V^L-V^R \right)\mathbf{y}$ is the force on the charge system inducing the charge current $\mathbf{I}=I\mathbf{y}$ along the transport direction $y$ as illustrated in Fig.~\ref{fig:ReciprocalProcesses}(c). Similarly, the spin force $\mathbf{X}_s^{L(R)}\equiv \boldsymbol{\mu}_s^{L(R)}/\hbar$ at the left (right) node induces the spin current $\mathbf{I}_s^{L(R)}$, Fig.~\ref{fig:ReciprocalProcesses}(a).  The magnetic system is described by the  magnetic free energy $F[\mathbf{M}]$ and the  macrospin $\mathbf{M}=VM_s\mathbf{m}$, where V is the volume of the ferromagnet, $M_s$ the absolute value of the magnetization, and $\mathbf{m}$ the unit direction vector. In this case, the effective field $\mathbf{X}_M\equiv -\partial_\mathbf{M} F[\mathbf{M}]$ induces the response $\dot{\mathbf{M}}$ in the magnetization.  
The relations between the rates and the thermodynamic forces are
\begin{equation}
\begin{pmatrix}
\dot{\mathbf{M}} \\
\mathbf{I}_s^L \\
\mathbf{I}_s^R \\
\mathbf{I} 
\end{pmatrix}
=
\begin{pmatrix}
\mathbf{L}_{MM} & \mathbf{L}_{Ms}^L & \mathbf{L}_{Ms}^R & \mathbf{L}_{Mc}  \\
\mathbf{L}_{sM}^L & \mathbf{L}_{ss}^{LL} & \mathbf{L}_{ss}^{LR} & \mathbf{L}_{sc}^L \\
\mathbf{L}_{sM}^R & \mathbf{L}_{ss}^{RL} & \mathbf{L}_{ss}^{RR} & \mathbf{L}_{sc}^R \\
\mathbf{L}_{cM} & \mathbf{L}_{cs}^L & \mathbf{L}_{cs}^R & \mathbf{L}_{cc}
\end{pmatrix}
\begin{pmatrix}
\mathbf{X}_M \\
\mathbf{X}_s^L \\
\mathbf{X}_s^R \\
\mathbf{X}_c 
\end{pmatrix}\,,
\label{MatrixEq}
\end{equation}
where $\mathbf{L}_{Ms}^{L(R)}$ characterizes the spin transfer torque arising from a spin accumulation in the left (right) node, $\mathbf{L}_{Mc}$ represents the torque from a charge potential, $\mathbf{L}_{MM}$ describes the magnetization dynamics for an isolated magnetic system according to the Landau-Lifshitz equation.

We identify the relations  
$\left[\mathbf{L}_{Ms}\left( \mathbf{m} \right)\right]_{ij}=\left[\mathbf{L}_{sM}\left( -\mathbf{m} \right)\right]_{ji}$ and $\left[\mathbf{L}_{Mc}\left( \mathbf{m} \right)\right]_{ij}=-\left[\mathbf{L}_{cM}\left( -\mathbf{m} \right)\right]_{ji}$. The $\left(\mathbf{L}_{sM}\right)_{ij}$ matrix describes spin pumping by a precessing magnetization. Its  reciprocal process is  spin transfer torque induced by a spin accumulation, as illustrated in Fig.~\ref{fig:ReciprocalProcesses}(a)-(b). Spin pumping is a well known phenomenon predicted theoretically~\cite{SpinPumping_Theory} and measured in experiments ~\cite{SpinPumping_Exp}. Charge pumping is described by $\left(\mathbf{L}_{cM}\right)_{ij}$ and caused by the spin-orbit interaction. Charge pumping is the reciprocal process of charge potential induced magnetization  dynamics, and is therefore a reciprocal manifestation of a charge current torque and non-vanishing coefficients $\left(\mathbf{L}_{Mc} \right)_{ij}$. This is illustrated in Fig.~\ref{fig:ReciprocalProcesses}(c)-(d).

The coefficients $\mathbf{L}_{Ms}^{L(R)}$ and $\mathbf{L}_{Mc}$ can be found from the out-of-equilibrium spin density in the ferromagnet, but such expressions are lengthy and cumbersome when spin-orbit coupling  is taken into account.  Instead, we consider the reciprocal processes, charge and spin pumping by magnetization precession, and compute the coefficients $\mathbf{L}_{sM}^{L(R)}$ and $\mathbf{L}_{cM}$. Parametric pumping can be found from the scattering matrix of the N$|$F$|$N system~\cite{SpinPumping_Theory} and using $\dot{M}_i= \left(\mathbf{L}_{MM} \right)_{ik} (\mathbf{X}_M)_k$ it reads: 
\begin{equation}
\hat{I}^{L(R)} =  \hat{Q}^{L(R)}_i \left(\mathbf{L}_{MM} \right)_{ik} (\mathbf{X}_M)_k\,,
\label{PumpingFormula}	 
\end{equation}
\begin{equation}						
\hat{Q}_{k}^{l}\equiv\frac{ie}{4\pi M_sV}\sum_{mnl'}\hat{s}_{mn,ll'}\partial_{m_k}\hat{s}^{\dagger}_{mn,ll'}+{\rm H.c.}\,,
\label{Qpump}
\end{equation}
where $\hat{s}_{mn,ll'}$ is the $2\times 2$ scattering matrix in spin-$1/2$ space from transverse mode $n$ in lead $l'$
to transverse mode $m$ in lead $l$. The current operator $\hat{I}^{L(R)}$ is in the  $2\times 2$ spin space which can be expanded in terms of charge and spin current contributions, $\hat{I}^{L(R)}= \hat{\sigma}_0 I^{L(R)}/2 + e\hat{\boldsymbol{\sigma}}\cdot\mathbf{I}_s^{L(R)}/\hbar$, where $\hat{\sigma}_0$ is the unit matrix and $\hat{\boldsymbol{\sigma}}$ is a vector of Pauli matrices.  This representation and Eq.~\eqref{PumpingFormula} give us the response coefficients 
$(\mathbf{L}_{cM})_{yk}= \text{Tr}[ \hat{Q}^{L}_i - \hat{Q}^{R}_i]\left(\mathbf{L}_{MM} \right)_{ik}/2$~\cite{comment2} and 
$(\mathbf{L}_{sM}^{L(R)})_{\alpha k}=\hbar\text{Tr}[\hat{\sigma}_{\alpha} \hat{Q}^{L(R)}_i ]\left(\mathbf{L}_{MM} \right)_{ik}/2e$, $\alpha\in \left\{x,y,z\right\}$.
Onsager reciprocal relations now imply that the magnetization dynamics induced by charge and spin currents is 
given by $\dot{M}_k= -[\mathbf{L}\left(-\mathbf{m}\right)_{cM}]_{yk}(V^L-V^R) + [\mathbf{L}(-\mathbf{m})_{sM}^{L(R)}]_{\alpha k}\mu_{s\alpha}^{L(R)}/ \hbar$.
Summarized, the generalized Landau-Lifshitz-Gilbert equation is
\begin{equation}
\dot{\mathbf{m}}= -\gamma\mathbf{m}\times\mathbf{H}_{\text{tot}} + \mathbf{m}\times \frac{\mathbf{\tilde{G}(m)}}{\gamma M_sV}\dot{\mathbf{m}} \label{EqOfMotion} \, , 
\end{equation}
where $\gamma$ is (minus) the gyromagnetic ratio, $\mathbf{H}_{\text{tot}}= \mathbf{H}_\text{eff}+\mathbf{H}^L_s+\mathbf{H}_s^R+\mathbf{H}_c$, and
\begin{align}
\mathbf{H}_c\left(\mathbf{m}\right) &\equiv  \text{Tr}[ \left(\mathbf{\hat{Q}}^{L}- \mathbf{\hat{Q}}^{R}\right) \left(-\mathbf{m}\right)] \left(V^L - V^R\right)/2, \label{EQHc} \\
\mathbf{H}_s^{L(R)}\left(\mathbf{m}\right) &\equiv  -\text{Tr}[\hat{\boldsymbol{\sigma}} \cdot\boldsymbol{\mu}_s^{L(R)} \mathbf{\hat{Q}}^{L(R)}\left(-\mathbf{m}\right) ]/2e \label{EQHs}\,. 
\end{align}
$\mathbf{H}_\text{eff}\equiv -\partial_{\mathbf{m}}F[\mathbf{m}]/VM_s$, and we have introduced the Gilbert damping tensor  in terms of the scattering matrix: $[\mathbf{\tilde{G}(m)}]_{ij}\equiv
\gamma^2\hbar\Re e\{ \text{Tr} [ ( \partial_{m_i} \hat{s} )(\partial_{m_j} \hat{s}^{\dagger }) ] \}/4\pi$ \cite{Brataas:prl08}. Eqs.~\eqref{EqOfMotion}, \eqref{EQHc}, and \eqref{EQHs} are our main results and capture general charge-current induced magnetization-torque effects induced by any spin-orbit coupling, interface torques as well as bulk torque effects. The conventional spin-transfer torques can only be included with normal leads, where the spin is conserved.   

To check our formalism, let us now consider a system without spin-flip processes and derive the spin transfer torque on $\mathbf{m}$ from a spin accumulation $\boldsymbol{\mu}_s^L$. For such a system, we expect our theory to be equivalent to magnetoelectronic circuit theory~\cite{Brataas:EurPhysJB01}. In absence of spin-orbit interactions the scattering matrix is written as 
$\hat{s}_{mn,ll'}= s_{mn,ll'}^{\uparrow}\hat{u}^{\uparrow} + s_{mn,ll'}^{\downarrow}\hat{u}^{\downarrow}$~\cite{SpinPumping_Theory} , where $\hat{u}^{\uparrow(\downarrow)}=(\hat{\sigma}_0 \pm \hat{\boldsymbol{\sigma}}\cdot\mathbf{m})/2$ is the matrix that projects out the spinor component for spin up (down) along the magnetization direction $\mathbf{m}$. Substituting this scattering matrix expression into Eq.~\eqref{EqOfMotion} gives the torque:
\begin{equation}
\boldsymbol{\tau} =
\frac{\gamma}{4\pi VM_s} \left(g_r^{\uparrow\downarrow}\mathbf{m}\times\boldsymbol{\mu}_s^L\times\mathbf{m}+g_i^{\uparrow\downarrow}\boldsymbol{\mu}_s^L\times\mathbf{m}  \right)\,.
\label{spinTorque}
\end{equation}
Here, we have defined $g^{\uparrow\downarrow}\equiv g_r^{\uparrow\downarrow}+ig_i^{\uparrow\downarrow}\equiv \sum_{mn}[ \delta_{mn} - r_{mn,LL}^{\uparrow}(r_{mn,LL}^{\downarrow})^{\ast}]$, and assumed the ferromagnet is wider than the transverse spin coherence length, implying $\sum_{mn} t_{mn,LR}^{\uparrow}(t_{mn,LR}^{\downarrow})^{\ast} \rightarrow 0$.
Eq.~\eqref{spinTorque} exactly agrees with magnetoelectronic circuit theory~\cite{Brataas:EurPhysJB01}, but is now derived from spin pumping via the Onsager relations.

%%%%%%%%%%%%%%%%%%%%%%%%%%%%%%%%%%%%%%%%%%%%%%%%%%%%%%%%%%%%%%%%%%%%%%%%%%%%%%%
%\section{Model}
%%%%%%%%%%%%%%%%%%%%%%%%%%%%%%%%%%%%%%%%%%%%%%%%%%%%%%%%%%%%%%%%%%%%%%%%%%%%%%%
We will in the following apply our theory to a disordered GaAs$|$(Ga,Mn)As$|$GaAs system, and
investigate the magnetization dynamics in the ferromagnetic (Ga,Mn)As layer induced 
by an unpolarized charge current.
We assume that the free-energy density depends on $\mathbf{m}$ as:
\begin{equation}
\mathcal{F}[\mathbf{m}]=K_{c1}\left( m_x^2m_y^2 + m_x^2m_z^2 + m_y^2m_z^2   \right)  + K_u m_z^2\,.
\label{freeEnergy}
\end{equation}
Here, $K_{c1}$ is the lowest-order cubic anisotropy constant~\cite{Modell}. We assume the system is grown on a GaAs substrate that induces compressive strain in the (Ga,Mn)As layer. Strain breaks the cubic symmetry. For growth direction $[001]$ this results in the unaxial anisotropy energy $K_um_z^2$~\cite{Modell}. We use the anisotropy constants $2K_{c1}/M_s=0.12$~T and $2K_{u}/M_s=0.35$~T \cite{Chiba:nature08}. 

To model the band structure of (Ga,Mn)As we use the following Hamiltonian \cite{Modell}:
\begin{equation}
  H = H_L + ~\mathbf{h(r)}\cdot \mathbf{J} + V(\mathbf{r}) + H_{\rm strain}.
\label{Hamiltonian}
\end{equation}
Here, $H_L$ is the $4\times4$ Luttinger Hamiltonian for zincblende semiconductors in the spherical approximation. $\mathbf{J}$ is a vector of $4 \!  \times \! 4$ spin matrices for $J \keq 3/2$ spins. The $\mathbf{h}\cdot\mathbf{J}$ term is a mean field approximation of the exchange interaction between the itinerant holes and the local magnetic moment of the Mn dopants. The exchange field $\mathbf{h}$ is antiparallel to the magnetization direction $\mathbf{m}$. $V(\mathbf{r})= \sum_i V_i\delta(\mathbf{r}-\mathbf{R}_i)$ is the impurity potential, where $\mathbf{R}_i$ is the position of impurity $i$,
and $V_i$ its scattering strength, which is randomly and uniformly distributed in the interval $[-V_0/2,V_0/2]$. Compressive strain in the (Ga,Mn)As layer induces the term $H_{\rm strain}$. For growth direction $[001]$, it is given by $\left(H_{\rm strain}\right)_{mn}=  -b\epsilon_{\text{ax}}\left(  J_z^2 - \mathbf{J}^2 / 3 \right)_{mn} - i C_4 \left(  J_y\partial_y -  J_x\partial_x \right)_{mn}$ \cite{Silver:prb92}, where $b= -1.7$~eV ~\cite{Silver:prb92}, $\epsilon_{\text{ax}}\equiv \left(\epsilon_{\text{zz}} - \epsilon_{\text{xx}} \right)= -0.005$ ~\cite{Modell,Silver:prb92},  $\epsilon_{n,n}$ are diagonal elements of the stress tensor, $C_4= 5$~eV~\AA~\cite{Silver:prb92},and $\delta_{mn}$ is the Kronecker delta. The GaAs leads are model by $H_L$, only.

We calculate the scattering matrix numerically using a stabilized transfer matrix method. We consider a discrete (Ga,Mn)As layer with transverse dimensions  $L_x= 19$~nm, $L_y= 50$~nm, and $L_z= 15$~nm. The lattice constant is $1$~nm, much less than the Fermi wavelength $\lambda_F \sim 10$~nm. The Fermi energy is $77$~meV when measured from the lowest subband in (Ga,Mn)As. To estimate a typical saturation value of the magnetization we use $M_s= 10|\gamma|\hbar x/a^3_{\rm GaAs}$~\cite{Modell}, with $x=0.05$ as the doping level, and where $a_{\rm GaAs}$ is the lattice constant for GaAs. The Luttinger parameters are $\gamma_1=7.0$ and $\gamma_2=2.5$, $|\mathbf{h}|=0.032$~eV, and $V_0=0.4$~eV for a diffusive system. Further details about the model and the numerical method can be found in Refs.~\cite{Modell}.

%%%%%%%%%%%%%%%%%%%%%%%%%%%%%%%%%%%%%%%%%%%%%%%%%%%%%%%%%%%%%%%%%%%%%%%%%%%%%%%
%\section{Results and discussion}
%%%%%%%%%%%%%%%%%%%%%%%%%%%%%%%%%%%%%%%%%%%%%%%%%%%%%%%%%%%%%%%%%%%%%%%%%%%%%%%
\begin{figure}[ht]
\centering
\includegraphics[scale=1.0]{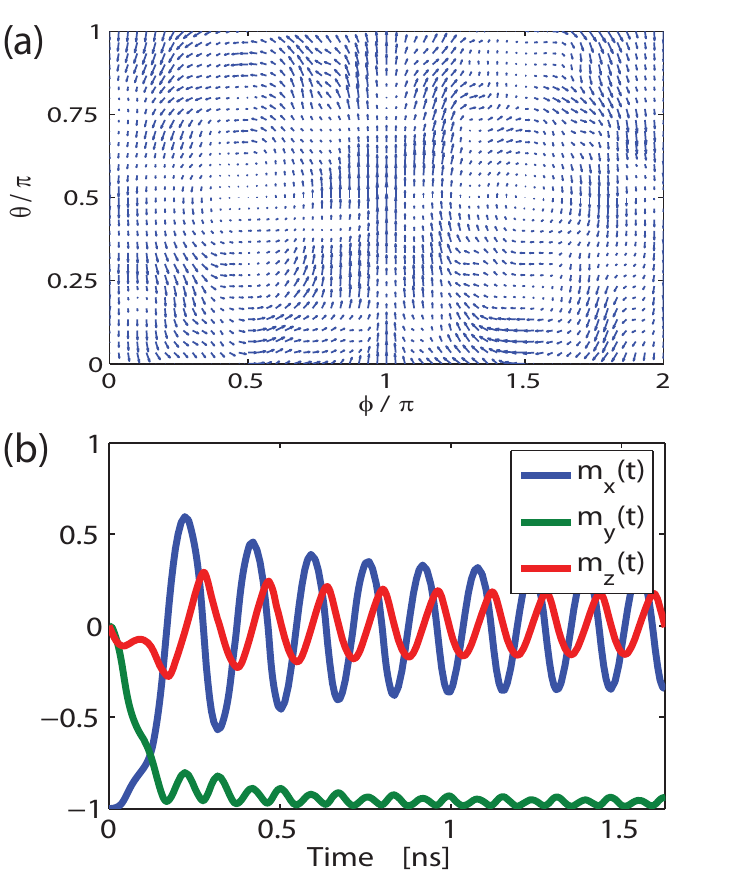}
\caption{(Color online) (a)  Vector field plot of $\left(\dot{\mathbf{m}}\right)_{\rm torque}\propto\mathbf{m}\times \mathbf{H}_c$  for a diffusive system with mean free path $l\sim 25$~nm. Here, $\theta$ and $\phi$ are respectively the polar and azimuth angles describing the local magnetization direction 
$\mathbf{m}=\left( \sin \theta  \cos \phi , \sin \theta  \sin \phi  , \cos \theta     \right)$.
(b) Solution of Eq.~\eqref{EqOfMotion} when applying a current density of $-1\times 10^{7}$~A/cm$^2$ along $y$.}
\label{fig:FieldPlots}
\end{figure}
We solve  Eq.~\eqref{EqOfMotion} for the magnetization direction $\mathbf{m}$ in (Ga,Mn)As  with $\mathbf{H}_c(\mathbf{m})$ and $\mathbf{\tilde{G}(m)}$ computed numerically from the scattering matrix. The scattering matrix is, for each magnetization direction, calculated for the (Ga,Mn)As region plus one lattice point into each of the GaAs layers. $\mathbf{H}_{\rm eff}$ is found from the free energy in Eq.~\eqref{freeEnergy}.   
The magnetization starts initially along the easy $x$ axis when an unpolarized current is applied to the system along the $y$ axis. We find for a diffusive system with mean free path $l\sim 25$~nm that current densities as low as $5\times 10^{6}$~A/cm$^2$ give magnetization switching between the $x$ and $y$ easy axis.

Fig.~\ref{fig:FieldPlots}(a)  shows the vector field $\left(\dot{\mathbf{m}}\right) _{\rm torque}\propto\mathbf{m}\times \mathbf{H}_c$ on the unit sphere $\mathbf{S}^2$ parameterized by the polar and azimuth angles $\theta$ and $\phi$. The strong circular flow around the $y$ axis, shows that $\mathbf{H}_c $  acts as an effective field along $y$.  This torque-field arises solely from the strain term of the Hamiltonian in Eq.~\eqref{Hamiltonian}~\cite{ Chernyshov:naturephys09, Garate:prb09,comment4}. In presence of an electric field along $y$,  $H_{\rm strain}\propto J_y k_y$ ($k_y$: Crystal momentum along $y$) tends to polarize the spin 3/2 holes along the $y$ axis. In combination with scattering at the GaAs$|$(Ga,Mn)As interfaces and impurities, this induces an out-of-equilibrium spin-density along y, giving the torque field shown in  Fig.~\ref{fig:FieldPlots}(a). 

Fig.~\ref{fig:FieldPlots}(b) shows a solution of  Eq.~\eqref{EqOfMotion} with the magnetization starting initially along $x$ when a current density of  $-1\times 10^{7}$~A/cm$^2$ is applied at time $t=0$ along $y$. The evolution of $\mathbf{m}$ can be understood as follows: The charge field $\mathbf{H}_c $  rotates $\mathbf{m}$ a small angel around the $y$ axis, giving a $m_z$ component to the magnetization. This induces an effective field component along $z$, arising from the large unaxial anisotropy energy $K_u  m_z^2$ in  Eq.~\eqref{freeEnergy}, that rotates $\mathbf{m}$ toward the y  easy axis where the magnetization equilibrates. The inverse process is obtained by applying a current along $x$.

Very recently, Ref.~\cite{Chernyshov:naturephys09} reported an experimental observation of charge current induced magnetization switching in (Ga,Mn)As  for current densities just below $10^{6}$~A/cm$^2$, while a theoretical study in Ref.~\cite{Garate:prb09} calculated critical currents of the order $10^{6}-10^{7}$~A/cm$^2$. This is in semi-quantitative  agreement with our results. In comparison, in metallic spin valve structures one typically observes critical spin currents of the order $10^6-10^8$~A/cm$^2$~\cite{SpinTransferTorqueReview}. Thus, charge current torque is not a negligible  effect compared to the conventional spin transfer torque, and will play an important role in modelling spintronic devices made out of ferromagnetic semiconductors and other systems with  strong spin-orbit coupling. 

%%%%%%%%%%%%%%%%%%%%%%%%%%%%%%%%%%%%%%%%%%%%%%%%%%%%%%%%%%%%%%%%%%%%%%%%%%%%%%%
%\section{Conclusion}
%%%%%%%%%%%%%%%%%%%%%%%%%%%%%%%%%%%%%%%%%%%%%%%%%%%%%%%%%%%%%%%%%%%%%%%%%%%%%%%
In conclusion, we have developed a general scattering matrix theory for magnetization 
dynamics that treats spin transfer torque and charge current torque 
induced by spin-orbit interaction. We apply our theory to a  layered 
GaAs$|$(Ga,Mn)As$|$GaAs system and show that unpolarized charge currents switches the  
magnetization direction for current densities as low as  $5\times 10^{6}$~A/cm$^2$, in agreement with recent experiments.

%%%%%%%%%%%%%%%%%%%%%%%%%%%%%%%%%%%%%%%%%%%%%%%%%%%%%%%%%%%%%%%%%%%%%%%%%%%%%%%
%\section{acknowledgment}
%\acknowledgments
%%%%%%%%%%%%%%%%%%%%%%%%%%%%%%%%%%%%%%%%%%%%%%%%%%%%%%%%%%%%%%%%%%%%%%%%%%%%%%%
We thank Anh Kiet Nguyen for developing the numerical transfer matrix code.
This work was supported in part by computing time through the Notur project and EC Contract IST-033749 "DynaMax".


\begin{thebibliography}{99}
\bibitem{comment2}
We assume that a possible Coulomb charging field in the ferromagnet repel/absorb charge equally into/from the two reservoirs so that the pumped current is given by $I=(I^L - I^R)/2$.
\bibitem{comment4}
In an inversion symmetric system the scattering matrix has the form $S= [r\  t ; t \  r]$, which implies $\mathbf{H}_c=0$. Thus,  $ H_{\rm strain}$ is the only term inducing a charge current torque in our model.
\end{thebibliography}
\end{document}